%% file: main.tex
\documentclass{article}

\usepackage{mathtools}
\usepackage{wrapfig}
\usepackage{tikz}

\usepackage[final]{neurips_2024}

\input{math_commands}
\usepackage[utf8]{inputenc} %
\usepackage[T1]{fontenc}    %
\usepackage{url}            %
\usepackage{booktabs}       %
\usepackage{amsfonts}       %
\usepackage{nicefrac}       %
\usepackage{microtype}      %
\usepackage{xcolor}         %

\definecolor{linkcolor}{rgb}{0,0.50,0.30} %
\definecolor{linkcolor}{rgb}{0,0.50,0.30} %
\usepackage{hyperref}
\hypersetup{
    colorlinks=true,
    allcolors=linkcolor
}

\definecolor{colorPhaseNet}{HTML}{fc8d59}
\definecolor{colorSeisLMLarge}{HTML}{006d2c}
\definecolor{colorSeisLMBase}{HTML}{74c476}
\definecolor{colorRandSeisLMBase}{HTML}{969696}
\definecolor{colorEarthquake}{HTML}{9e0142}
\definecolor{colorNoise}{HTML}{91bfdb}

\newcommand{\markerNoise}[1][colorNoise,fill=colorNoise]{%
  \tikz{\draw[#1] (0.75ex,0.75ex) circle (0.8ex);}%
}

\newcommand{\markerEarthquake}[1][colorEarthquake,fill=colorEarthquake]{%
  \tikz{\draw[#1] (0,0) -- (1.5ex,0) -- (0.75ex,1.3ex) -- cycle;}%
}

\newcommand{\markerPhaseNet}[1][colorPhaseNet,fill=colorPhaseNet]{%
  \tikz{\draw[#1] (0.75ex,0.75ex) circle (0.8ex);}%
}

\newcommand{\markerSeisLMLarge}[1][colorSeisLMLarge,fill=colorSeisLMLarge]{%
  \tikz{\draw[#1] (0,0) rectangle (1.4ex,1.4ex);}%
}

\newcommand{\markerRandSeisLMBase}[1][colorRandSeisLMBase,fill=colorRandSeisLMBase]{%
  \tikz{\draw[#1] (0,0) rectangle (1.4ex,1.4ex);}%
}

\newcommand{\markerSeisLMBase}[1][colorSeisLMBase,fill=colorSeisLMBase]{%
  \tikz{\draw[#1] (0,0) -- (1.5ex,0) -- (0.75ex,1.3ex) -- cycle;}%
}

\title{SeisLM: a Foundation Model for Seismic Waveforms}

\author{%
 Tianlin Liu \\
  University of Basel \\
  \And
  Jannes Münchmeyer \\
  Université Grenoble Alpes,\\ Université Savoie Mont Blanc, \\ CNRS, IRD, Université Gustave Eiffel, ISTerre
  \And
  Laura Laurenti \\
  Sapienza University of Rome 
  \And
  Chris Marone \\
  Sapienza University of Rome\\ Pennsylvania State University 
  \And
  Maarten V. de Hoop \\
  Rice University 
  \And
  Ivan Dokmanić\thanks{Contact authors: Tianlin Liu <\texttt{t.liu@unibas.ch}> and Ivan Dokmanić <\texttt{ivan.dokmanic@unibas.ch}>.} \\
  University of Basel \\
}

\begin{document}

\maketitle

\begin{abstract}
We introduce the Seismic Language Model (SeisLM), a foundational model designed to analyze seismic waveforms---signals generated by Earth's vibrations such as the ones originating from earthquakes. SeisLM is pretrained on a large collection of open-source seismic datasets using a self-supervised contrastive loss, akin to BERT in language modeling. This approach allows the model to learn general seismic waveform patterns from unlabeled data without being tied to specific downstream tasks.  When fine-tuned, SeisLM excels in seismological tasks like event detection, phase-picking, onset time regression, and foreshock--aftershock classification. The code has been made publicly available on \url{https://github.com/liutianlin0121/seisLM}.
\end{abstract}

\section{Introduction}

Seismology is a data-centric field that often sees significant progress through improvements in data quality and quantity \citep{Havskov2010routine,Zhou2014practical}. Today, the field benefits from an extensive collection of seismic recordings gathered over years by networks of thousands of stations worldwide \citep{Hafner2001southern, Mousavi2019stanford, Quinteros2021geofon, Michelini2021instance, Cole2023mlaapde, Niksejel2024obstransformer, Chen2024txed, Zhong2024deep}. Over the last decades, millions of these recordings have been manually inspected and labeled by domain experts. This wealth of data and labels has fueled the rise of machine-learning models, which automate the analysis of these expanding seismic records.
A growing body of models, including convolutional networks \citep{Ross2018generalized, Zhu2018phasenet, Woollam2019convolutional, Mousavi2019cred}, recurrent networks \citep{Soto2021deepphasepick, Yoma2022end}, and transformers \citep{Mousavi2020earthquake, Li2024SeisT, munchmeyer2021transformer} have been applied to seismic data analysis, particularly in tasks like earthquake detection and characterization.

Despite these advances, most current machine-learning models in seismology still depend on \emph{labeled, task-specific datasets}, not making use of more than a petabyte of openly available unlabeled waveforms.
This mirrors the early stages of machine learning in fields like computer vision and natural language processing, where models were initially trained on similarly specialized datasets such as MNIST \citep{LeCun1998gradient}, CIFAR \citep{Krizhevsky2009learning}, Sentiment140 \citep{Go2009twitter}, and IMDB dataset \citep{Maas2011learning}. Yet, these task-specific models eventually gave way to general-purpose foundation models, trained on a wealth of unlabled data, which are capable of handling a broader range of tasks with minimal fine-tuning.
Exemplars of open-weight foundation models include BERT \citep{Devlin2019bert}, GPT-2 \citep{Radford2019language}, and Llama \citep{Touvron2023llama, Touvron2023llama2, Dubey2024llama} for text processing, Wav2Vec2 \citep{Baevski2020wav2vec} and Hubert \citep{Hsu2021hubert} for speech understanding, and CLIP \citep{Radford2021learning} and MAE \citep{He2022masked} for vision modeling.
These foundation models rely on \emph{self-supervised learning} from unlabeled data, allowing them to scale up training samples and learn features without being tied to specific tasks.

In this work, we introduce the Seismic Language Model (SeisLM), a self-supervised model for analyzing single-station seismic waveforms.  SeisLM uses a standard encoder-only transformer architecture, similar to Wav2Vec2 and BERT. Our results demonstrate that this model, when pretrained on worldwide earthquake activity records, extracts generalizable features that effectively address various downstream tasks, nearly always surpassing models tailored for specific tasks. The main contributions of the paper are summarized below:

\begin{itemize}
    \item We introduce a self-supervised foundation model for seismic waveforms. To our knowledge, it represents the first application of self-supervised learning on unlabeled seismic waveforms.
    
    \item We demonstrate that the model's self-supervised features, although not trained on any labeled samples, display clear and interpretable characteristics. Specifically, the model groups waveform features into noise and earthquake clusters.
    
    \item We show that the self-supervised model generalizes well to a wide array of downstream tasks. When compared with supervised baselines, the advantage of pretraining--finetuning is particularly noticeable when the downstream tasks have limited labeled data.
    
\end{itemize}

\section{Background and related work}

\paragraph{Supervised-learning models for seismic tasks.}
The efforts of using supervised machine learning to automate seismic waveform analysis stretch back several decades. We briefly review a non-exhaustive selection of neural network approaches. Early methods used shallow multilayer perceptrons (MLPs) to classify seismic waveforms \citep{Enescu1996seismic, Baevski2020wav2vec, Dai1997application, Zhao1999artificial, Gentili2006automatic}. Starting from 2010s, 1D convolutional neural networks (ConvNets) have been prevalent in seismic applications due to their efficiency and flexibility in handling variable-length input. For instance, the Generalized Phase Detection model \citep{Ross2018generalized} uses a 1D convolutional network for phase classification tasks. Inspired by the U-Net \citep{Ronneberger2015unet}, a convolutional network originally designed for 2D image segmentation, \citet{Zhu2018phasenet, Woollam2019convolutional} used similar architectures in 1D for onset and phase picking tasks. \citet{Mousavi2019cred} proposed a residual convolutional network for earthquake detection, drawing on ideas from residual networks used in image classification \citep{He2016deep}. In addition to ConvNets, recurrent networks (RNNs) have also been applied to seismic tasks. These networks include DeepPhasePick \citep{Soto2021deepphasepick}, which handles event detection and phase picking. Finally, the recent success of transformers and their self-attention mechanisms \citep{Vaswani2017attention} has inspired their use in seismic analysis. The Earthquake Transformer \citep{Mousavi2020earthquake} combines recurrent networks and self-attention mechanisms for joint event detection, phase detection, and onset picking. While Earthquake Transformer is a Transformer--CNN--RNN hybrid approach, Seismogram transformer \citep{Li2024SeisT} shows that a plain transformer can be used to solve different earthquake-monitoring tasks when coupled with different head modules.

\paragraph{Unsupervised learning models for seismic tasks.}
Unsupervised machine learning has been used to uncover patterns in unlabeled seismic data, primarily through clustering and visualization. \citet{Esposito2008unsupervised} cluster volcanic event waveforms to explore the link between active volcanic vents and their explosive waveforms. \citet{Yoon2015earthquake} group waveforms with similar features in a database, then use a search method to identify those resembling earthquake signals. \citet{Mousavi2019unsupervised} use convolutional autoencoders to cluster and differentiate hypocentral distances and first-motion polarities. \citet{Seydoux2020clustering} combine scattering networks with a Gaussian mixture model to cluster seismic signal segments, demonstrating applications in blind detection and recovery of repeating precursory seismicity.

\paragraph{Foundation models for seismic tasks and their relationships to our work.}
There exist a few foundation models for seismic applications, although they differ from our approach in several aspects.
\citet{Sheng2023seismic} proposed a foundation model for \emph{seismic imagery data}, which are visual representations of the Earth's subsurface structures. These images are generated by seismic waves reflecting off rock boundaries, capturing the differences in physical properties between various geological layers.
In contrast, our work focuses on seismic waveforms, which are time-series data. In this regard, the closest related models are \citet{Si2024seisclip} and \citet{Li2024SeisT}, which also handle seismic waveforms. Both, however, rely on labeled datasets for pretraining. Specifically, \citet{Si2024seisclip} uses event annotations, such as phase and source information, for a contrastive approach. \citet{Li2024SeisT} uses a \emph{supervised pretraining} method, training a single model for various classification and regression tasks, including earthquake detection and phase picking, using labeled data.
Our approach is distinct in that we use only \emph{unlabeled waveforms} for pretraining. 
This is motivated by the consideration that unlabeled waveforms are much more accessible and abundant than labeled ones. To our knowledge, SeisLM is the first foundation model self-supervisedly trained on unlabeled seismic waveforms.

\section{Seismic Language Model}

\begin{figure}[ht!]
\centering
\includegraphics[width=1.0 \linewidth]{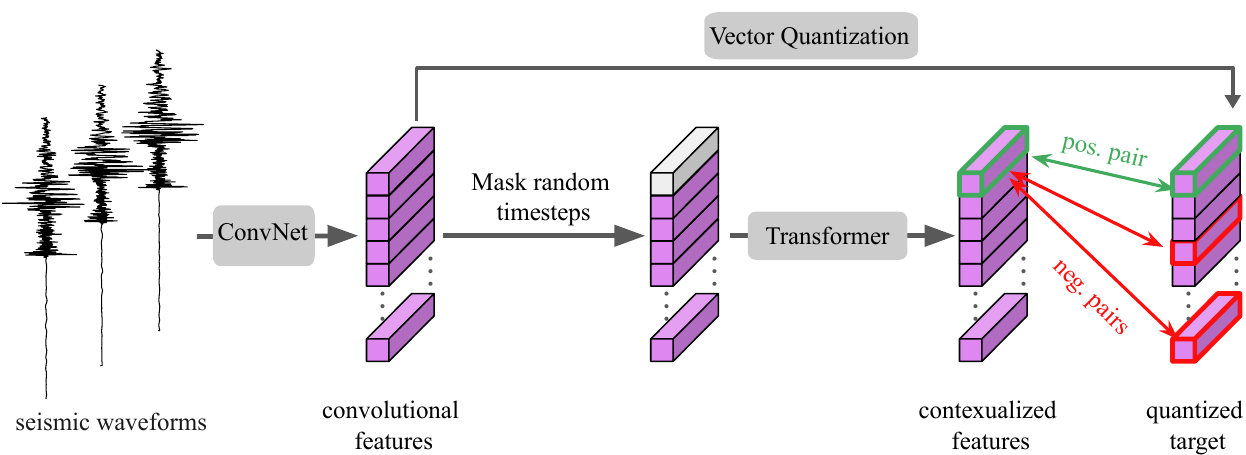}
\caption{\textbf{Illustration of the self-supervised learning of Seismic Language Model (SeisLM).} A ConvNet encodes raw 3-channel seismic waveforms from a single station into a feature sequence. The model then follows two paths. In the lower path, we apply random masking to these waveform features before passing them to a transformer. The transformer aims to reconstruct aspects of the masked convolutional features. In the upper path, we prepare the reconstruction targets: continuous-valued convolutional features are discretized into a sequence of vectors with a finite vocabulary size through vector quantization (VQ; \citealp{Van2017neural, Razavi2019generating}). This overall model closely resembles Wav2vec2 \citep{Baevski2020wav2vec} for audio self-supervised learning. \label{fig:seislm-model}}
\end{figure}

Our language model is an encoder-only transformer that focuses on the task of predicting features of masked timesteps. This model architecture is standard, closely following Wav2Vec2 \citep{Baevski2020wav2vec} for speech signal modeling and BERT \citep{Devlin2019bert} for text modeling. In Fig.~\ref{fig:seislm-model}, we show a general overview of the model, which consists of a ConvNet, a quantizer, and a transformer. We now explain the role of each module and defer their detailed hyperparameters to Section~\ref{sec:experiment}.

\subsection{SeisLM architecture \label{subsec:seislm-architecture}}

\paragraph{Model input.}
The input to the model are raw seismic waveforms, which are a sequence of vectors $(\vx_1, \ldots, \vx_{T})$; each sample $\vx_t \in \RR^3$ has three channels that correspond to ground motion recorded by a single seismometer for three orthogonal axes: East--West, North--South, and Up--Down; this format is standard in seismic data. Most seismic datasets use a sampling rate of 100 Hz or of the same order of magnitude (see Table~\ref{tab:pretraining-dataset}); we thus use waveforms at 100 Hz for consistency and resample the waveform to 100Hz in case the original sampling rate differs. We standardize each channel of a waveform by subtracting the channel mean and dividing by the channel standard deviation. 

\paragraph{ConvNet encoder.} 
The raw waveforms $(\vx_1, \ldots, \vx_T)$ first undergo a 1D ConvNet, yielding convolutional features $(\vv_1, \ldots, \vv_{L})$ with $\vv_t \in \RR^{d_v}$.
The purpose of the 1D ConvNet is twofold: (i) filter the raw waveform and lift the 3-dimensional waveform signals to a higher dimension ($d_v > 3$), and (ii) downsample the sequence of the raw waveform in length ($L < T$), so that self-attention layers can be applied to this shorter sequence with lower computational complexity.

\paragraph{Transformer encoder.} 
The convolutional features are then fed into a sequence of transformer blocks \citep{Vaswani2017attention} after masking and position embedding. The masking part replaces convolutional features at random timesteps by a fixed embedding vector (details in Section~\ref{subsec:pretraining-setup}). For position embedding, as in \citet{Baevski2020wav2vec}, we apply a 1D group convolutional layer \citep{Krizhevsky2012imagenet} with a large kernel to obtain relative positional embedding, and then sum the output with the masked features. The position-embedded masked features are then fed to the transformer. The transformer is the heart of the model, as its self-attention mechanism \citep{Vaswani2017attention} captures contextual information.  We write the transformer output as  $(\va_1, \ldots, \va_L)$ with $\va_t \in \RR^{d_q}$.

\paragraph{Quantization.} During pretraining, the transformer encoder aims to reconstruct the unmasked convolutional seismic features from their masked corruptions. We use \emph{quantized} convolutional features as the reconstruction targets: Given an input $\vv_t \in \RR^{d_v}$ of the raw waveform, the quantization module \citep{Jegou2010product} intuitively retrieves the nearest neighbor of $\vv_t$ over a finite codebook $\Qcal \coloneqq \{\vq_{(1)}, \ldots, \vq_{(n_q)}\} \subset \RR^{d_q}$ and use the resulting vector as the target; the parenthesized indices here refer to the enumeration of the code vectors, which differs from the unparenthesized ones used to denote timesteps.
Using quantized waveforms as the target proved more effective than non-quantized waveforms in previous speech self-supervised learning research \citep{Baevski2020wav2vec, Baevski2019vq}. \citet{Baevski2020wav2vec} suggested that quantization reduces specific artifacts, such as speaker and background noise, which simplifies the reconstruction task and prevents the model from fitting on irrelevant details. To obtain the quantized vectors, a quantization module $Q: \RR^{d_v} \to \Qcal$ is applied to the feature vector at each timestep independently with $\vq_t \coloneqq Q(\vv_t)$.  To parameterize the quantization function $Q$, we follow \citet{Jegou2010product} and use learnable matrices $\mW \in \RR^{n_q \times d_v}$ to compute
\begin{align}
    [\vz_1, \ldots, \vz_L] & = \textrm{LayerNorm} \Big([\vv_1, \ldots, \vv_L] \Big) \\
     i_t & \coloneqq \argmax \big(\mW \vz_t \big)  \in \{1, \ldots, n_q\},~~\textrm{for all}~~ t \in [L] \label{eq:argmax-selection}\\
    \vq_t & \coloneqq \vq_{(i_t)} \in \Qcal \subset \RR^{d_q}.
\end{align}
Here, $\argmax \big(\mW \vz_t \big)$ indicates the entry to the largest value of the vector $\mW \vz_t$. Since $\argmax$ is not differentiable, in practice, we use the Gumbel-Softmax trick \citep{Jang2017categorical} as a differentiable relaxation of the argmax in the forward pass of the model.
Furthermore, following \citet{Baevski2020wav2vec}, we introduce multiple codebooks, identify one codeword from each of the codebook, and then concatenate them. This concatenation approach increases the number of possible quantization vectors at the expense of more parameters; for example, if we use two codebooks, each with $n_q$ codewords, then the total possible quantization vectors is $n_q^2$.

\section{Training}

To pretrain the SeisLM, we use a masked reconstruction objective similar to masked language modeling in BERT \citep{Devlin2019bert} and masked audio modeling in Wav2vec2 \citep{Baevski2020wav2vec}. For each masked time step, the pretraining goal is to identify the correct quantized latent representation from a candidate set. After the pretraining, the model is finetuned on labeled samples.

\subsection{Pretraining setup \label{subsec:pretraining-setup}}

\paragraph{Masking.} A portion of the convolutional features $(\vv_1, \ldots, \vv_L)$ is randomly replaced by a shared trainable feature vector during each forward pass of pertaining. To select the masking indices, similar to \citet{Baevski2020wav2vec}, we uniformly sample $6.5 \%$ of all time-steps to be starting indices and mask the subsequent 10 time-steps. 

\paragraph{Contrastive loss.}

We pretrain SeisLM with a standard contrastive objective: At each timestep $t$, we encourage the transformer output $\va_t$ to positively correlate with the quantized feature vector $\vq_t$ of the same timesteps, and negatively correlate with $K$ quantized feature vectors sampled from other timesteps of the same input sequence. Denoting these $K$ negative examples at each timestep $t$ by $\Ncal_{t} \coloneqq \{\vn_t^{1}, \ldots, \vn_{t}^{K}\} \subset \{\vq_1, \ldots, \vq_L\}$, we let the contrastive loss of each time $t$ be
\begin{equation} \label{eq:contrastive-loss}
L(\va_t, \vq_t, \Ncal_t) \coloneqq - \log \frac{\exp \Big [ \mathrm{sim} (\va_t, \vq_t) / \kappa \Big]  }{ \exp \Big [\mathrm{sim} (\va_t, \vq_t) / \kappa  \Big] + \sum_{\vn \in \Ncal_t} \exp \Big [\mathrm{sim} (\va_t, \vn) / \kappa \Big ]}.    
\end{equation}
where $\kappa > 0$ is a fixed temperature.

While $L(\vq_t, \Qcal_t)$ in \eqref{eq:contrastive-loss} is the main loss used for masked pretraining, we add auxiliary losses to encourage the codebook vectors in $\Qcal$ to be less redundant; this is achieved with an entropy regularization as in \citet{Baevski2020wav2vec}.

\paragraph{Codevector diversity loss.}

Optimizing the quantization module faces the common issue of underutilized codebooks \citep{Dieleman2018challenge, Lancucki2020robust, Dhariwal2020jukebox, Mentzer2023finite}: codewords may remain unused. To address this, following prior works \citep{Baevski2020wav2vec, Dieleman2018challenge}, we use a diversity loss to encourage the uniform use of codebook vector. Concretely, let $\{\vv_1, \ldots, \vv_{BL}\}$ be a batch of $B$ covolutional waveveform sequences, each with length $L$; we let $\{\vp_1, \ldots, \vp_{BL}\}$ be the softmax probabilities of codevector assignment: $\vp_j \coloneqq \textrm{softmax}(\mW \vz_j) \in \RR^{n_q}$, which is a differentiable relaxation of the hard assignment in \eqref{eq:argmax-selection}. The average of these codevector assignment probabilities, $\overline{\vp} \coloneqq \frac{1}{BL} \sum_{j=1}^{BL} \vp_j \in \RR^{n_q}$ is another probability vector that describes the average usage of all codevector. The diversity loss is defined as $\frac{1}{n_q} \langle \overline{\vp}, \log \overline{\vp} \rangle.$ However, this diversity loss can itself lead to numerical instability if its strength is not carefully tuned. Our experience shows that this instability is in part due to the \emph{highly unbalanced codebook usage at initialization}. This imbalance triggers a large diversity loss at the outset, leading to substantial initial optimization updates as the model tries to correct it. In Appendix~\ref{appendix:model-details}, we propose a simple way to initialize the model such that the diversity loss remains small. During pretraining, we combine the diversity loss with the contrastive loss. The balance between them is controlled by a hyperparameter.

\subsection{Finetuning setup}
To finetune pretrained models to a downstream, labeled dataset task, we add a randomly initialized shallow network to process the output of SeisLM. Since SeisLM down-samples waveforms through its convolutional layers, the transformer output has a shorter length than the raw input. Thus, for sequence-labeling tasks that predict each timestep at the original frequency, we use linear interpolation followed by convolutional layers to upsample the latent representation; more details are in Appendix~\ref{appendix:experiments}. During the finetuning, we simply train the parameters of both the SeisLM and the task head. We are aware of prior work that freezes some parts of the pretrained model or uses a scheduler to gradually unfreeze the pretrained model \citep{Baevski2020wav2vec} during finetuning; however, these more involved approaches did not bring consistent improvement in our finetuning experiments.

\section{Experiment \label{sec:experiment}}

\subsection{Pretraining experiments}

\begin{table*}[th]
 \centering
\setlength{\tabcolsep}{3pt}
\footnotesize
\begin{tabular}{l c c  c c c c} \toprule
& \bf Traces  & \bf Region  & \bf Tr. length & \bf Sampling rate [Hz] & \bf Type\\
\midrule
ETHZ & 36,743   & Switzerland  & variable & 100 - 500 & Regional\\
INSTANCE & 1,291,537  & Italy  & 120~s & 100 & Regional\\
Iquique & 13,400  & Northern Chile &  variable & 100 & Regional\\
STEAD & 1,265,657  &  global &  60~s & 100 & Regional\\
GEOFON & 275,274  &   global &   variable & 20 - 200 & Teleseismic\\
MLAAPDE & 1,905,887 & global & 120 & 40 & Teleseismic\\
PNW & 183,909 & Pacific Northwest & 150~s & 100 & Regional\\
OBST2024 & 60,394  & global & 60~s & 100 & Regional, submarine\\ \bottomrule
\end{tabular}
\label{dataset_table}
 \caption{Overview of the pretraining datasets from SeisBench \citep{Woollam2022seisbench}. While waveforms from these datasets come with various labels such as phase labels (e.g., P-phase vs S-phase), we only use the raw, unannotated data in the training fold for pretraining. \label{tab:pretraining-dataset}}
\end{table*}

\paragraph{Pretraining data.} For the pretraining dataset, we combine waveforms from eight seismic datasets, accessed through the SeisBench \citep{Woollam2022seisbench} framework, into a unifying dataset. The eight datasets are ETHZ \citep{SED1983,SED2005,SED2008,Alparray2014,CERN2016}, INSTANCE \citep{Michelini2021instance}, Iquique \citep{Woollam2019convolutional}, STEAD \citep{Mousavi2019stanford}, GEOFON \citep{Quinteros2021geofon}, MLAAPDE \citep{Cole2023mlaapde}, PNW \citep{Ni2023curated}, and OBST2024 \citep{Niksejel2024obstransformer}. These datasets consist of preselected waveform snippets, encompassing examples of earthquakes, noise, and exotic signals such as explosions and landslides. Due to this preselection, the prevalence of earthquake signals in this data is substantially higher than on a randomly recorded seismic trace. An overview of these datasets is provided in Table~\ref{tab:pretraining-dataset}. These datasets cover examples from different world regions, different event-to-station distances, and a wide magnitude range. We randomly sample 30s windows from the traces.

\begin{figure}[t]
\centering
\includegraphics[width=0.8 \linewidth]{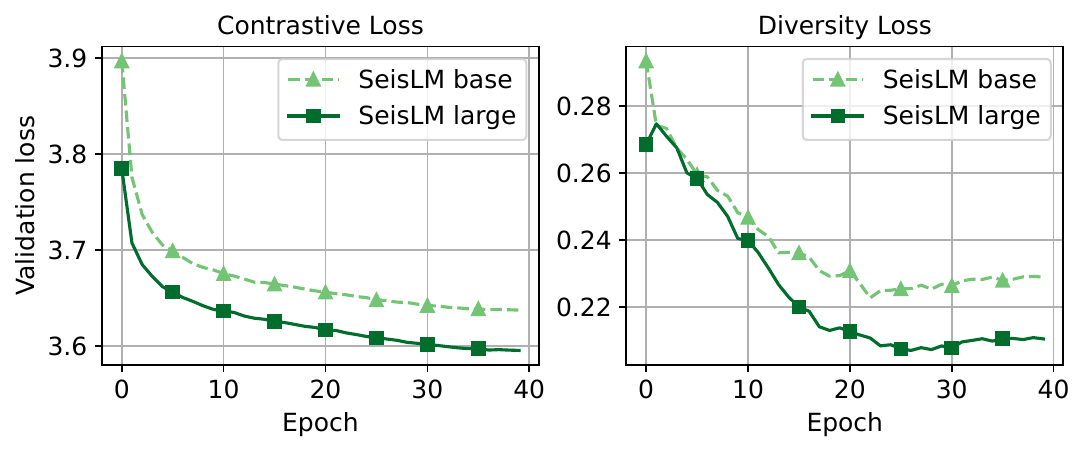}
\caption{\textbf{Pretraining loss of SeisLM. \label{fig:pretrain-loss}}}
\end{figure}

\paragraph{Model and training hyperparameters.} 
We briefly outline the hyperparameters used in pretraining and provide full details in Appendix~\ref{appendix:experiments}. We pretrained two variants of models: \emph{SeisLM-base} and \emph{SeisLM-large}. They share the same ConvNet and quantization configurations but \emph{SeisLM-large} uses a larger transformer module than \emph{SeisLM-base}: SeisLM-base includes 6 transformer blocks, while SeisLM-large has 12. The SeisLM-base contains 11.4 million parameters, while SeisLM-large contains 90.7 million parameters. We trained our model with the Adam optimizer \citep{Kingma2015adam} for 40 epochs. We trained SeisLM-base on four A100-40G GPUs and trained SeisLM-large on four A100-80G GPUs. the pretraining of SeisLM-base and SeisLM-large takes approximately 5 and 8 days, respectively. Figure~\ref{fig:pretrain-loss} plots the validation losses of two SeisLM models during pretraining.

\begin{figure}[h]
\centering
\includegraphics[width=0.8 \linewidth]{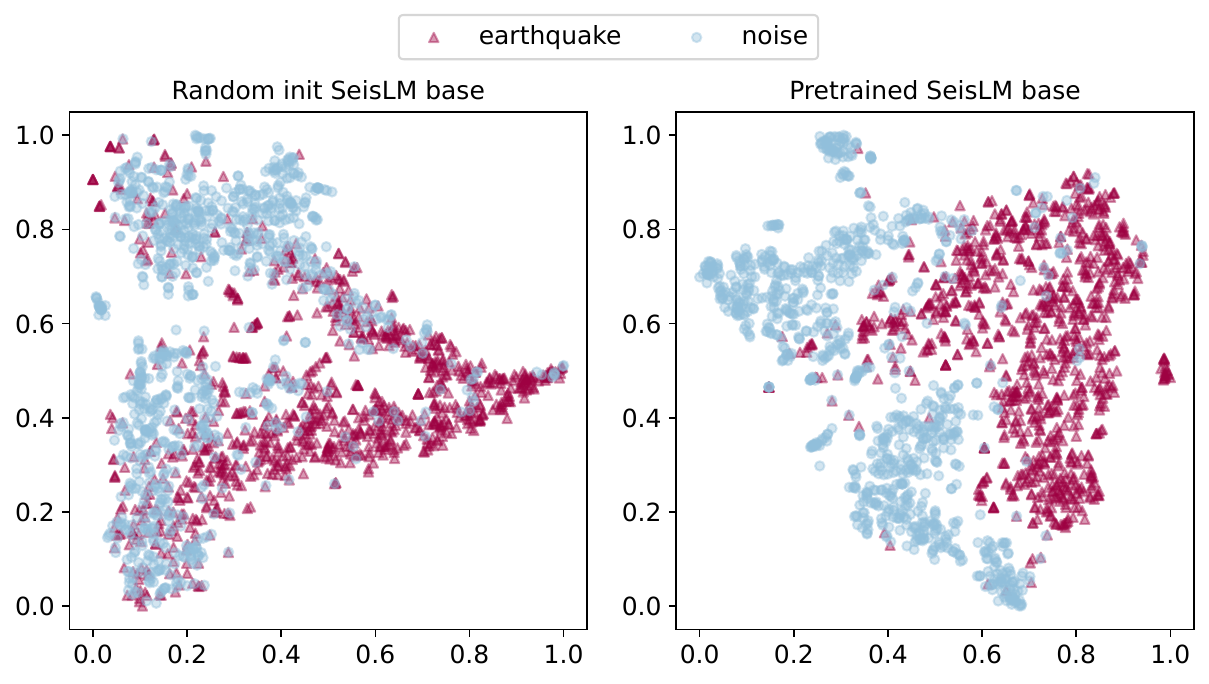}
\caption{\textbf{t-SNE embeddings of SeisLM features.} Compared to a randomly initialized SeisLM-base (left panel), a self-supervised SeisLM-base (right panel) separate the embeddings of earthquake and noise waveforms more effectively. \label{fig:tsne}}
\end{figure}

\paragraph{Visualizing learned features through dimensionality reduction.}
Does the reduction of pretraining loss, shown in Figure~\ref{fig:pretrain-loss}, mean that the model learns useful features from the data? As a sanity check, we run a simple dimensionality reduction experiment. This experiment visualizes whether the pretrained SeisLM, without fitting on any labeled data, could reasonably separate noise and earthquake traces. We collect 1000 noise traces and 1000 earthquake traces from the INSTANCE dataset and input them into SeisLM. For each trace, we average the features from the last layer of SeisLM along the time axis, producing one embedding vector per trace. This process is akin to the bag-of-words model in natural language processing. We apply t-SNE \citep{Maaten2008tsne} to non-linearly reduce the dimensionality of the trace embeddings to $2$, to facilitate visualization (Figure~\ref{fig:tsne}). The results indicate that, with randomly initialized weights, the SeisLM embeddings of noise ($\markerNoise$) and earthquake ($\markerEarthquake$) traces heavily overlap (left panel of Figure~\ref{fig:tsne}); however, after self-supervised pretraining, the separation between the embeddings of noise and earthquake traces gets greatly improved (right panel of Figure~\ref{fig:tsne}). 
We emphasize again the embeddings are learned without using any label; they are colored using labels in Figure~\ref{fig:tsne} for probing purposes.

\subsection{Finetuning on phase-picking tasks \label{sec:finetuning-phasepick}}
We now test whether self-supervised SeisLMs transfer effectively to downstream seismic tasks. Among the many potential downstream tasks, detecting and determining seismic phase types and their onset time are arguably the most fundamental ones; these tasks are typically jointly referred to as \emph{phase-picking} tasks. 
More specifically, seismic phase onset time is the moment of seismic waves emitted by a source, such as an earthquake, reach a seismic instrument; we usually observe two main phase types of seismic waves, the faster longitudinal P waves and the slower S waves. The results of seismic phase picking form the basis of many subsequent seismological workflows, in particular, earthquake detection through phase association \citep{zhu2022earthquake,munchmeyer2024pyocto}, source characterization \citep{Bormann2012new} or seismic travel-time tomography \citep{nolet1987seismic}.
All of these steps are integral for accurate and precise seismic hazard assessment.

For a quantitative analysis, we consider the three evaluation tasks defined in the large-scale benchmark by \citet{Munchmeyer2022picker}:
\begin{enumerate}
\item \textbf{Event detection}: Given a window of a seismic waveform, determine if it contains an event.
\item \textbf{Phase identification}: Given a window containing exactly one phase arrival, determine if it is a P or an S phase.
\item \textbf{Onset regression}: Given a window containing exactly one phase arrival of the known type (P or S), determine the onset time.
\end{enumerate}
We show event detection and phase identification results in the main text, and place the onset regression result in Appendix~\ref{appendix:experiments}.

\textbf{Setup of the baseline models and SeisLMs.} In the benchmark study of \citet{Munchmeyer2022picker}, PhaseNet \citep{Zhu2018phasenet} achieves the best overall performance for the three phasepicking tasks described above. We, therefore, use PhaseNet as a baseline, with the same PhaseNet hyperparameters as in \citet{Munchmeyer2022picker}.
Note that PhaseNet solves the three-way phase-picking task: for each sample, PhaseNet outputs a 3-dimensional probability vector corresponding to the noise probability, P-phase probability, and the S-phase probability \citep{Zhu2018phasenet, munchmeyer2021transformer}.
For a head-to-head comparison, we follow this joint-training approach to finetune SeisLM. We add two convolutional layers on top of the pre-trained SeisLM with a Softmax activation function in the end, so that it outputs a 3-dimensional probability vector at each timestep, just like the PhaseNet. More details of the finetuning hyperparameters are in Appendix~\ref{appendix:experiments}.
For both models, we use 1 minus the noise probability for the event detection.
We use the ratio of the peak of the P and S as predictions for the phase identification task.
We use the peak position of the relevant phase prediction for the onset regression task.

\paragraph{Finetuning dataset.} We use three labeled phase-picking datasets from Seisbench for finetuning \citep{Woollam2022seisbench, Munchmeyer2022picker}: ETHZ, GEOFON, and STEAD. These datasets reflect different data availability scenarios: ETHZ contains 22k training traces (low data), GEOFON provides 161k traces (medium data), and STEAD offers more than 1 million traces (abundant data). To evaluate model performance across various sample sizes, we divide each dataset into fractions, ranging from 5\% to 100\%. This allows us to test the models with varying amounts of labeled data. We hypothesize that pretrained models to perform much better than randomly initialized networks in low-data scenarios. In abundant-data scenarios, we anticipate that randomly initialized networks will also perform well, but pretraining should not hinder performance; therefore, we include the large STEAD dataset to stress test the pretrained model.

\begin{figure}[h]
\centering
\includegraphics[width=1.0 \linewidth]{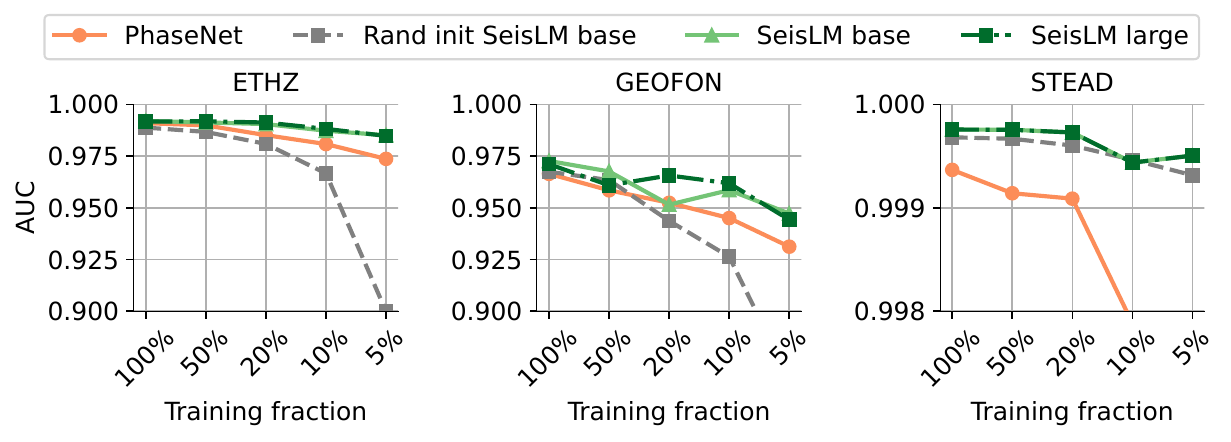}
\caption{\textbf{Performance of models on the event detection task.} Each panel indicates a finetuning dataset. The $x$ axis indicate the fraction of training dataset; the $y$ axis shows the AUC metric: it represents the area under the curve that plots the true positive rate against the false positive rate at various threshold levels for a binary classification task.
\label{fig:all_datasets_event_detection}}
\end{figure}

\paragraph{Event detection.} Figure~\ref{fig:all_datasets_event_detection} illustrates the event identification results across three datasets. When comparing event detection accuracy at various fractions of the training dataset, pretrained SeisLM models ($\markerSeisLMBase$, $\markerSeisLMLarge$) consistently outperformed PhaseNet ($\markerPhaseNet$). The advantage of SeisLM is especially pronounced with a limited number of labeled samples, such as when using just $5\%$ of the training data.
However, the difference in performance between SeisLM-base ($\markerSeisLMBase$) and SeisLM-large ($\markerSeisLMLarge$) is minimal, presumably because this event detection is relatively simple task.
Additionally, we compared a SeisLM model fine-tuned from pretrained weights ($\markerSeisLMBase$, $\markerSeisLMLarge$) with a SeisLM-base model initialized with random weights ($\markerRandSeisLMBase$). The results show that pretraining benefits performance, particularly when labels are scarce. When there is sufficient labeled data, such as the case of STEAD dataset, then a randomly initialized SeisLM can perform reasonably well.

\begin{figure}[ht!]
\centering
\includegraphics[width=1.0 \linewidth]{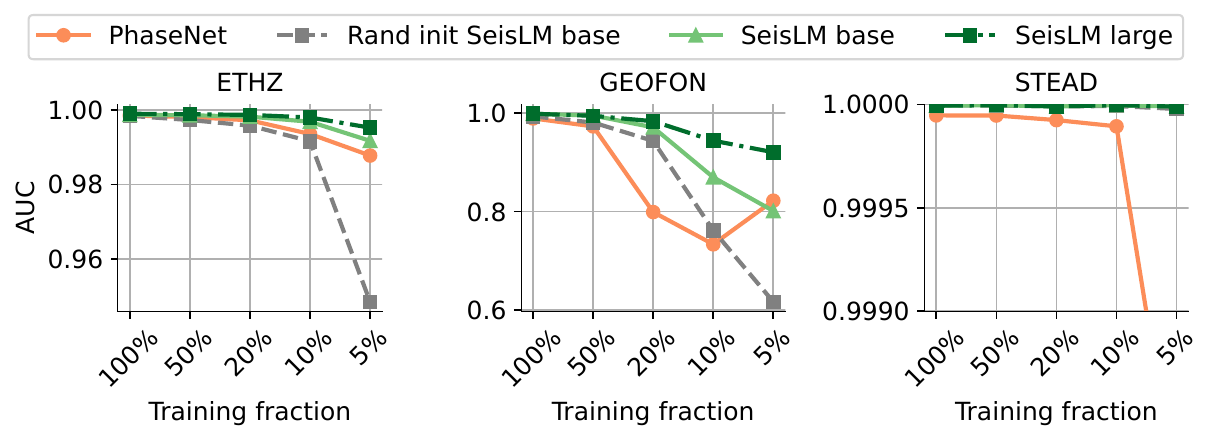}
\caption{\textbf{Performance of models on the phase identification task.} \label{fig:all_datasets_auc_phase_identification}}
\end{figure}

\paragraph{Phase identification.} Figure~\ref{fig:all_datasets_auc_phase_identification} displays phase identification results across the same three datasets. As with event detection, pretrained SeisLM models ($\markerSeisLMBase$, $\markerSeisLMLarge$) generally deliver higher accuracy than models trained from scratch ($\markerPhaseNet$, $\markerRandSeisLMBase$), with the gap widening in low-data scenarios. Additionally, SeisLM-large ($\markerSeisLMLarge$) surpasses SeisLM-base ($\markerSeisLMBase$) in this task. When using a substantial amount of data from the largest STEAD dataset, all SeisLM models---whether randomly initialized or pretrained---perform the task near perfect.

\subsection{Finetuning on foreshock--aftershock classification tasks}
A major challenge in seismology is detecting subtle changes in seismic recordings before and after earthquakes. Gaining insights to these subtle changes can offer early warnings of impending hazards. 
Previous research has impressively shown that machine learning models can be trained to identify foreshock and aftershock seismic waves \citep{Laurenti2024probing}. Specifically, \citet{Laurenti2024probing} classified waveform signals into different categories based on the time relative to the 2016 M6.5 Norcia mainshock in Italy. We apply SeisLM to tackle the same task.

\begin{figure}[t]
\centering
\includegraphics[width=1.0 \linewidth]{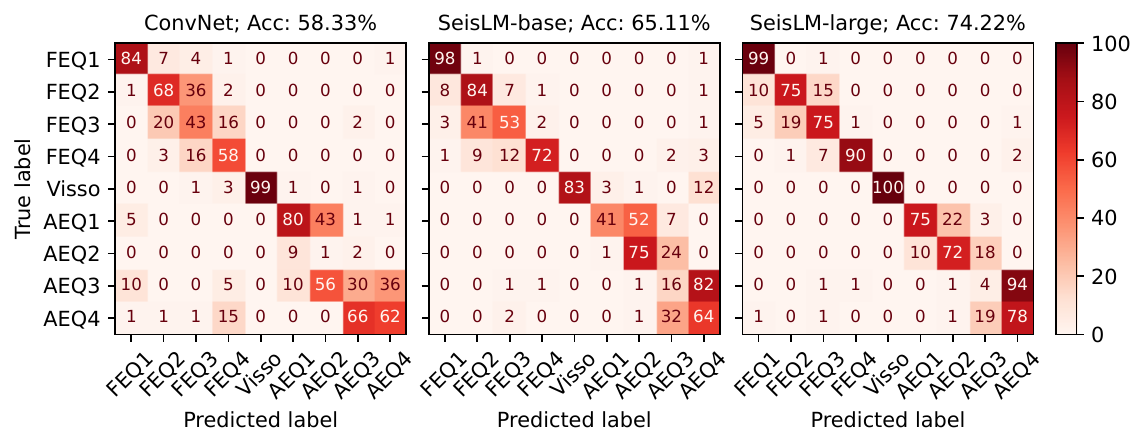}
\caption{\textbf{Confusion matrices of models evaluated on the test fold of the foreshock-–aftershock classification dataset.} The $x$-axis represents the predicted labels, and the $y$-axis represents the true labels. The values in the matrices indicate the percentage of predicted samples. The event classes are ordered by time.\label{fig:shock_result}}
\end{figure}

\paragraph{Data and model.} Following the exact dataset setting of \citet[Section 3.1.1]{Laurenti2024probing}, we focused on the waveform recordings from the NRCA station. The foreshock, mainshock, and aftershock events are categorized into nine classes, ranging from FEQ1 (earliest foreshocks), to Visso (the main shock), and finally to AEQ4 (latest aftershocks). These classes are displayed as the labels of the $x$ and $y$ of Figure~\ref{fig:shock_result}. We use the 7-layer ConvNet from \citet[Section 8.2.1]{Laurenti2024probing} as our baseline model. To fine-tune SeisLM, we add convolutional layers on top of its transformer block; these convolutional layers are followed by global average pooling and a linear head. See Appendix~\ref{appendix:experiments} for more details.

\paragraph{Results.} Figure~\ref{fig:shock_result} displays the confusion matrices on the test-fold of the foreshock--aftershock dataset. SeisLM's fine-tuning (middle and right panels) improves accuracy over the ConvNet baseline (left panel). Furthermore, reassuringly, the confusion matrices show that SeisLM's errors often occur in temporal proximity---e.g., misclassifying FEQ2 traces as FEQ3 traces and vice versa. Overall, our results provide further support to the hypothesis in \citet{Laurenti2024probing}: fault or source properties before and after a major earthquake show detectable changes that can be identified in seismic recordings.

\section{Discussion}
Foundation models for seismic waveforms are in their early stages, and important insights are still missing. Take model scaling, for example. In text modeling, researchers have investigated the optimal model size and token count for training transformers within a fixed compute budget, most notably through the Chinchilla scaling law \citep{Hoffmann2022training}. We currently lack comparable insights for seismic tasks.
Despite this, SeisLM shows the promise of self-supervised learning on unlabeled seismic waveforms---the same strategy behind many seminal foundation models in vision and language modelling. This self-supervised approach enables the pre-trained model to excel in downstream tasks, often surpassing task-specific baselines. It becomes especially helpful when labeled data for downstream tasks is scarce.

The early stage of seismic foundation model research is in contrast with their potential for immense impact. Indeed, earthquakes rank among the most dangerous natural hazards, and even small advances in early warning and hazard assessment could substantially improve safety and reduce economic damage. Leveraging the petabytes of existing seismic data---and likely exponentially more from emerging technologies \citep{shearer2023earthquake, zhan2020distributed}---self-supervised learning methods applied to vast amounts of unlabeled seismic data may significantly improve seismic data analysis. With the introduction of SeisLM, we have taken a step in this direction.

\paragraph{Acknowledgment}
Tianlin Liu and Ivan Dokmanić were supported by the European Research Council Starting Grant 852821---SWING. Numerical experiments were partly performed at the sciCORE (\url{http://scicore.unibas.ch/}) scientific computing center at the
University of Basel. 
Jannes Münchmeyer is funded by the European Union under the grant agreement n°101104996---DECODE.

\clearpage
\bibliographystyle{mybst.bst}
\bibliography{project.bib}

\appendix

\section{Details of the model \label{appendix:model-details}}

\paragraph{Quantization.}

\begin{figure}%
  \begin{center}
    \includegraphics[width=0.48\textwidth]{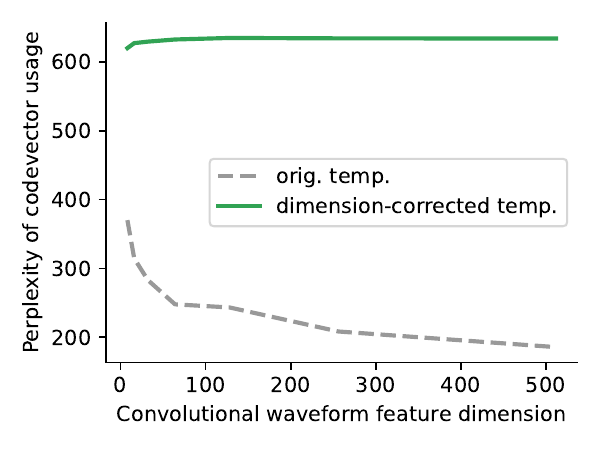}
  \end{center}
  \caption{The influence of the standard temperature $\tau$ and dimension-corrected temperature $\tau/\sqrt{n_q}$ in a randomly initialized Gumbel quantizer. When the convolutional feature dimension $d_v$ (x-axis) increases, the perplexity of codevector (y-axis) increases in the case of standard temperature (grey curve); this indicates uneven usage of codebook vectors. With the dimensionality correction, the perplexity stays roughly constant (green curve). \label{fig:dimension-corrected-temp}}
\end{figure}

We show that this phenomenon of uneven usage of randomly initialized codebooks can be easily understood. During training, the forward pass of the Gumbel-quantizer computes

\begin{align}
    \vp_t & \coloneqq \textrm{softmax} \big[(\mW \vz_t + \vn)/\tau \big] \in \RR^{n_q}, \quad \textrm{with}~\vn_j \overset{\textrm{iid}}{\sim} \textrm{Gumbel}(0, 1) \textrm{~for all~} j \in [n_q], \\
    i_t & \sim \text{Categorical}(\vp_t), \quad i_t \in \{1, \ldots, n_q\} 
\end{align}
where $\tau$ is a temperature. At initialization, the entries of the weight projection matrix $\mW$ are typically drawn from a Normal distribution\footnote{As in the implementation of Gumbel quantizer of \href{https://github.com/facebookresearch/fairseq/blob/920a548ca770fb1a951f7f4289b4d3a0c1bc226f/fairseq/modules/gumbel_vector_quantizer.py\#L79C34-L79C45}{Fairseq} and \href{https://github.com/huggingface/transformers/blob/a26de151390f5cb029b2e39231c00ad4303b4347/src/transformers/models/wav2vec2/modeling_wav2vec2.py\#L1346}{Hugging Face transformer}}. Assume that the convolutional feature $\vz \in \RR^{d_v}$ follows a normal distribution. In this case, the entries of $(\mW \vz + \vn)/\tau$ follow a zero-mean Gaussian distribution with variance proportional to $d_v$, the dimension of convolutional features. Given that $d_v$ is typically in the order of hundreds, the variance is in the same order, leading to nearly one-hot vectors after the softmax. This makes the categorical sampling nearly deterministic and less exploratory for codevectors. Additionally, since larger models often use greater codevector dimensions $d_v$, larger models more prone to this problem. We illustrate this in Figure~\ref{fig:dimension-corrected-temp}. As a simple fix, we re-parametrize the temperature $\tau$ as $\tau \coloneqq \tau' \sqrt{n_q}$. This re-parametrization breaks the link between the convolutional feature dimension $d_v$ and its impact on uneven codevector usage at initialization.

\section{Experimental details \label{appendix:experiments}}

\subsection{Pretraining experiments}

\paragraph{Model hyperparameters}
We pretrained two variants of models: \emph{SeisLM-base} and \emph{SeisLM-large}. They share the same ConvNet and quantization configurations but \emph{SeisLM-large} uses a larger transformer module than \emph{SeisLM-base}.
For the ConvNet module, each model uses two convolutional layers with 256 channels, a kernel size of 3, and a stride of 2. In the vector quantization module, each model uses two groups of code vectors, each containing 320 vectors.
Furthermore, each model's position embedding component (placed at the start of the transformer module) uses a grouped convolutional layer \citep{Krizhevsky2012imagenet} with a kernel size of 128 and 16 groups.
In the rest of the transformer module, SeisLM-base includes 6 pre-norm transformer blocks, while SeisLM-large has 12. Unlike the standard transformer block, the pre-norm version applies layer normalization before the self-attention and feedforward layers. This modification often leads to more stable training \citep{Baevski2018adaptive,Nguyen2019transformers,Xiong2020layer}. Each transformer block employs a 12-headed self-attention layer and a residual 2-layer MLP with 3072 hidden units. The number of output units of the MLP is 240 for SeisLM-base and 768 for SeisLM-large. With these settings, SeisLM-base contains 11.4 million parameters, while SeisLM-large contains 90.7 million parameters.

\paragraph{Training hyperparameters}
For the contrastive loss, we randomly sample $K=100$ quantization vectors from the convolutional feature sequences as negative examples, with a temperature $\kappa = 0.1$ in \eqref{eq:contrastive-loss}. We trained our model with the Adam optimizer \citep{Kingma2015adam} for 40 epochs. 
We traind SeisLM-base with a global batch size of 112 on four A100-40G GPUs, and trained SeisLM-large with a global batch size of 192 on four A100-80G GPUs.
The learning rate scheduler uses cosine annealing with a linear warmup. The maximum learning rate is 5e-4 for SeisLM-base and 1e-3 for SeisLM-large, with the same warmup fraction of 20\%. During training, we decreased the Gumbel temperature from 2.0 to 0.5. We did not apply dropout, drop layers, or weight decay during pretraining. We trained SeisLM-base with 16-bit precision and SeisLM-large with 32-bit precision. With these settings, the pretraining of SeisLM-base and SeisLM-large takes approximately 5 and 8 days, respectively. Figure~\ref{fig:pretrain-loss} plots the validation losses of two SeisLM models during pretraining.

\subsection{Phase-picking experiments}

\paragraph{Hyperparameters of the finetuned SeisLM.}
Since Pretrained SeisLM down-samples waveforms through its convolutional layers, the transformer's output is shorter than the raw input. For phase-picking tasks, we upsample the latent representation to match the input length using linear interpolation. We then concatenate this upsampled representation with the raw waveforms and apply two convolutional layers to fit the labels. Specifically, we use two convolutional layers with a kernel width of 3, stride of 1, and GELU activations. These layers maintain the number of channels in the transformer features. For fine-tuning SeisLM-base, we use 240 + 3 convolutional filters, and for SeisLM-large, we use 768 + 3 filters. We also apply dropout with a rate of 0.2 after each convolutional layer.

\begin{figure}[ht!]
\centering
\includegraphics[width=1.0 \linewidth]{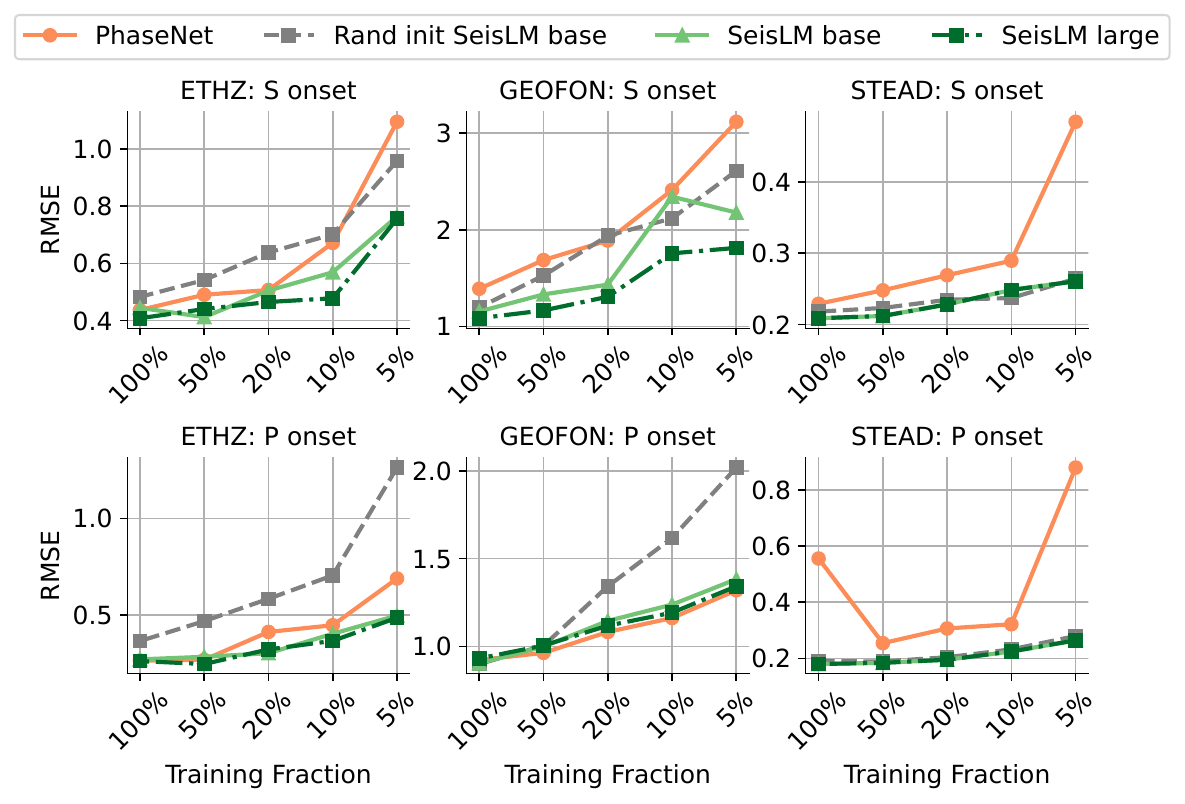}
\caption{\textbf{Onset regression.} \label{fig:all_datasets_onset_regression}}
\end{figure}

\paragraph{Onset regression.} Figure~\ref{fig:all_datasets_onset_regression} displays the onset regression result, which recapitulated our findings on the two phasepicking tasks above. Pretrained SeisLM ($\markerSeisLMBase$, $\markerSeisLMLarge$) generally achieves lower onset regression than train-from-scratch baselines ($\markerPhaseNet$, $\markerRandSeisLMBase$).

\subsection{Foreshock--aftershock experiments}

\paragraph{Hyperparameters of the finetuned SeisLM.}
For foreshock--aftershock tasks, we add a 4-layer convolutional network on top of pretrained SeisLM. These convolutional layers have a kernel width of 3, stride of 2, and GELU activations, and they maintain the number of channels in the transformer features. A global average pooling layer and a linear head follow the convolutional layers, turning the features into a vector of $9$ classes.

\end{document}

%% file: math_commands.tex
\usepackage{amsmath,amsfonts,bm}

\def\1{\bm{1}}

\def\va{{\bm{a}}}

\def\vn{{\bm{n}}}

\def\vp{{\bm{p}}}
\def\vq{{\bm{q}}}

\def\vv{{\bm{v}}}

\def\vx{{\bm{x}}}

\def\vz{{\bm{z}}}

\def\mW{{\bm{W}}}

\DeclareMathAlphabet{\mathsfit}{\encodingdefault}{\sfdefault}{m}{sl}
\SetMathAlphabet{\mathsfit}{bold}{\encodingdefault}{\sfdefault}{bx}{n}

\def\Ncal{{\mathcal{N}}}

\def\Qcal{{\mathcal{Q}}}

\def\RR{{\mathbb{R}}}

\DeclareMathOperator*{\argmax}{arg\,max}

\usepackage{hyperref} 
    \hypersetup{
    	colorlinks = true,
    	linkcolor = blue,
    	anchorcolor = blue,
    	citecolor = blue,
    	filecolor = blue,
    	urlcolor = blue
    }